\documentclass[10pt,twocolumn,twoside]{IEEEtran}
\IEEEoverridecommandlockouts
\usepackage{cite}
\usepackage{amsmath,amssymb,amsfonts}
\usepackage{algorithmic}
\usepackage{textcomp}
\usepackage{xcolor}
\usepackage{subfigure}
\usepackage{caption}
\usepackage{booktabs}
\usepackage{algorithm}
\usepackage{multirow}
\usepackage{graphicx}
\usepackage{comment}
\usepackage{makecell}
\usepackage{thmtools, thm-restate}
\newtheorem{theorem}{Theorem}
\def\BibTeX{{\rm B\kern-.05em{\sc i\kern-.025em b}\kern-.08em
    T\kern-.1667em\lower.7ex\hbox{E}\kern-.125emX}}
\DeclareMathAlphabet{\mathscr}{OT1}{pzc}{m}{it}

\begin{document}
\newtheorem{definition}{\it Definition}
\newtheorem{lemma}{\it Lemma}
\newtheorem{corollary}{\it Corollary}
\newtheorem{remark}{\it Remark}
\newtheorem{example}{\it Example}
\newtheorem{case}{\bf Case Study}
\newtheorem{assumption}{\it Assumption}
\newtheorem{property}{\it Property}
\newtheorem{proposition}{\it Proposition}

\newcommand{\hP}[1]{{\boldsymbol h}_{{#1}{\bullet}}}
\newcommand{\hS}[1]{{\boldsymbol h}_{{\bullet}{#1}}}

\newcommand{\ba}{\boldsymbol{a}}
\newcommand{\baq}{\overline{q}}
\newcommand{\bA}{\boldsymbol{A}}
\newcommand{\bb}{\boldsymbol{b}}
\newcommand{\bB}{\boldsymbol{B}}
\newcommand{\bc}{\boldsymbol{c}}
\newcommand{\bcO}{\boldsymbol{\cal O}}
\newcommand{\bh}{\boldsymbol{h}}
\newcommand{\bH}{\boldsymbol{H}}
\newcommand{\bl}{\boldsymbol{l}}
\newcommand{\bm}{\boldsymbol{m}}
\newcommand{\bn}{\boldsymbol{n}}
\newcommand{\bo}{\boldsymbol{o}}
\newcommand{\bO}{\boldsymbol{O}}
\newcommand{\bp}{\boldsymbol{p}}
\newcommand{\bq}{\boldsymbol{q}}
\newcommand{\bR}{\boldsymbol{R}}
\newcommand{\bs}{\boldsymbol{s}}
\newcommand{\bS}{\boldsymbol{S}}
\newcommand{\bT}{\boldsymbol{T}}
\newcommand{\bw}{\boldsymbol{w}}

\newcommand{\balpha}{\boldsymbol{\alpha}}
\newcommand{\bbeta}{\boldsymbol{\beta}}
\newcommand{\bOmega}{\boldsymbol{\Omega}}
\newcommand{\bTheta}{\boldsymbol{\Theta}}
\newcommand{\bphi}{\boldsymbol{\phi}}
\newcommand{\btheta}{\boldsymbol{\theta}}
\newcommand{\bvarpi}{\boldsymbol{\varpi}}
\newcommand{\bpi}{\boldsymbol{\pi}}
\newcommand{\bpsi}{\boldsymbol{\psi}}
\newcommand{\bxi}{\boldsymbol{\xi}}
\newcommand{\bx}{\boldsymbol{x}}
\newcommand{\by}{\boldsymbol{y}}

\newcommand{\cA}{{\cal A}}
\newcommand{\bcA}{\boldsymbol{\cal A}}
\newcommand{\cB}{{\cal B}}
\newcommand{\cE}{{\cal E}}
\newcommand{\cG}{{\cal G}}
\newcommand{\cH}{{\cal H}}
\newcommand{\bcH}{\boldsymbol {\cal H}}
\newcommand{\cK}{{\cal K}}
\newcommand{\cM}{{\cal M}}
\newcommand{\cN}{{\cal N}}
\newcommand{\cO}{{\cal O}}
\newcommand{\cP}{{\cal P}}
\newcommand{\cR}{{\cal R}}
\newcommand{\cS}{{\cal S}}
\newcommand{\dcS}{\ddot{{\cal S}}}
\newcommand{\ds}{\ddot{{s}}}
\newcommand{\cT}{{\cal T}}
\newcommand{\cU}{{\cal U}}
\newcommand{\cV}{{\cal V}}
\newcommand{\cW}{{\cal W}}
\newcommand{\cX}{{\cal X}}
\newcommand{\cY}{{\cal Y}}
\newcommand{\cZ}{{\cal Z}}

\newcommand{\hcx}{{\hat {\cal x}}}
\newcommand{\hcX}{{\hat {\cal X}}}
\newcommand{\hw}{{\hat w}}
\newcommand{\hcW}{{\hat {\cal W}}}

\newcommand{\wt}[1]{\widetilde{#1}}

\newcommand{\mA}{\mathbb{A}}
\newcommand{\mE}{\mathbb{E}}
\newcommand{\mG}{\mathbb{G}}
\newcommand{\mR}{\mathbb{R}}
\newcommand{\mS}{\mathbb{S}}
\newcommand{\mU}{\mathbb{U}}
\newcommand{\mV}{\mathbb{V}}
\newcommand{\mW}{\mathbb{W}}

\newcommand{\uq}{\underline{q}}
\newcommand{\ubq}{\underline{\boldsymbol q}}

\newcommand{\red}[1]{\textcolor[rgb]{1,0,0}{#1}}
\newcommand{\gre}[1]{\textcolor[rgb]{0,1,0}{#1}}
\newcommand{\blu}[1]{\textcolor[rgb]{0,0,0}{#1}}

\title{Rate-Distortion Theory for Strategic Semantic Communication}

\author{\IEEEauthorblockA{Yong~Xiao\IEEEauthorrefmark{1}\IEEEauthorrefmark{3}, Xu Zhang\IEEEauthorrefmark{1}, Yingyu Li\IEEEauthorrefmark{1}, Guangming~Shi\IEEEauthorrefmark{2}\IEEEauthorrefmark{3}, Tamer Ba\c{s}ar\IEEEauthorrefmark{4} \\
\IEEEauthorblockA{\IEEEauthorrefmark{1}School of Elect. Inform. \& Commun., Huazhong Univ. of Science \& Technology, China}\\
\IEEEauthorblockA{\IEEEauthorrefmark{2}School of Artificial Intelligence, Xidian University, Xi'an, China}\\
\IEEEauthorblockA{\IEEEauthorrefmark{3}Pengcheng National Laboratory (Guangzhou), Guangzhou, China}\\
\IEEEauthorblockA{\IEEEauthorrefmark{4}Depart of Elect. \& Comp. Eng., University of Illinois Urbana-Champaign, Urbana, Illinois, USA}\\
}\thanks{%
Corresponding author: Guangming~Shi

*This paper is accepted at IEEE Information Theory Workshop, Mumbai, India, November 2022.}
}

\maketitle

\begin{abstract}
This paper analyzes the fundamental limit of the strategic semantic communication problem in which a  transmitter obtains a limited number of indirect observations of an intrinsic semantic information source and can then influence the receiver's decoding by sending a limited number of messages over an imperfect channel. The transmitter and the receiver can have different distortion measures and can make rational decisions about their encoding and decoding strategies, respectively. The decoder can also have some side information (e.g., background knowledge and/or information obtained from previous communications) about the semantic source to assist its interpretation of  the semantic information. We focus particularly on the case that the transmitter can commit to an encoding strategy and study the impact of the strategic decision making on the rate distortion of semantic communication. Three equilibrium solution concepts including the optimal Stackelberg equilibrium, robust Stackelberg equilibrium, as well as Nash equilibrium are studied and compared. The optimal encoding and decoding strategy profiles under various equilibrium solutions are derived. We prove that committing to an encoding strategy cannot always bring benefit to the encoder. We provide a feasible condition under which committing to an encoding strategy can always reduce the distortion of semantic communication. We consider an example with a dictionary-based semantic information source to verify our observation.    
\end{abstract}


\vspace{-0.2in}
\section{Introduction}
Utilizing semantic information during the communication process has been a long-term vision in information theory. Shannon in his  seminal work published in 1948 has already noticed that most of the communication ``messages have meaning"\cite{shannon1948mathematical}. However, in the development of the mathematical theory of communication, the ``semantic aspect of messages" has been intentionally ignored because the semantic meaning of a message can be correlated with “certain physical and conceptual entities” and therefore making the meaning as part of the communication may affect the generality of the theory\cite{shannon1948mathematical}. The concept of semantic communication problem has been formally introduced in 1949 by Weaver where the problems addressed by Shannon theory have been coined into the technical problem of communication and the {\em semantic communication problems} have been defined as ``the problems that are concerned with the identity, or satisfactorily close approximation, in the interpretation of meaning by the receiver, as compared with the intended meaning of the sender"\cite{weaver1949recent}. Since then, many efforts have been made to extend the Shannon theory to model the communication of semantic information\cite{Carnap1952SemanticInfo,Bao2011SemanticComm,Liu2021SemanticInformation,shi2020semantic}. For example, the authors in \cite{Carnap1952SemanticInfo} have replaced the statistical probability measure of information in Shannon theory with the logical probability measure between constants and predicates to analyze semantic information carried by sentences in a given language system. In \cite{Bao2011SemanticComm}, the authors observed that one of the unique features of the semantic communication problems is that the background knowledge can be available at both the transmitter and the  receiver to assist the interpretation of semantic meaning.
Some other works also suggested that environmental information as well as background knowledge can some time play a key role in improving the performance of semantic communication\cite{Juba2008UniverSemanticComm}. 
Another result suggested that, unlike traditional source of information, the semantic aspects of information may consist of intrinsic features or states and may not always be directly observable by the encoder\cite{Liu2021SemanticInformation}. 

Despite of the above progress, there are limitations to directly extending the classic Shannon theory to investigate the semantic communication problem. For example, in \cite{Carnap1952SemanticInfo}, the authors have observed a contradiction, commonly referred to as the Bar-Hillel-Carnap (BHC) paradox, when applying the Shannon's measure of information to quantify the volume of semantic message. In the BHC paradox, it is observed that self-contradictory or semantically-false sentences should have less value or no  meaningful information. These sentences, however, often contain much more information than fact or common-knowledge-based sentences when measured using entropy-based metric due to their rarity. In \cite{Floridi2010InformationBook,Floridi2012SemanticInfo,StanfordEncyclopedia2022SemanticInfo}, the authors have studied the semantic information from the philosophical perspective and observed that the semantic information should not focus only on accurately reproducing the content-agnostic data, but the interpretation of the key idea at the receiver. 

Motivated by the above observations, we investigate, in this paper, semantic communication problems from a strategic communication perspective\cite{Akyol2015StrategicCompression,Akyol2016OnTheRoleStrategicComm,Akyol2017StrategicComm}. We argue that combining game theory, a mathematical tool to model strategic interaction among rational agents\cite{Nisan2007AlgorithmGameTheory}, with information theory may have the potential to address the limitations of the classical Shannon theory and obtain a more comprehensive solution profile for semantic communication problems due to the following reasons. First, game theory models interactions and potential influence between rational users who focus on maximizing their rewards based on mutually agreed rules and common beliefs. A transmitter  will have less incentive to send meaningless information that cannot bring any reward, e.g., less or no influence, to the receiver, and hence could have the potential to avoid the BHC paradox. Second, different from the classic Shannon theory focusing on maximizing the total volume of data transported from one point to another, strategic decision making allows game playing agents such as transmitters and receivers to achieve more flexible objectives, e.g., coordinated interpretation of the key idea of semantic message and/or implicitly influencing semantic interpretation, especially in capacity limited scenarios\cite{Akyol2015StrategicCompression,Akyol2016OnTheRoleStrategicComm}. For example, in \cite{Treust2019Persuasion,Treust2020StrategicCommSideInfo,Treust2021ISITStrategicCommSideInfo,kamenica2011bayesian}, the authors  extended the Bayesian persuasion problem into information theoretic setting in which the transmitter, instead of transporting all the detailed data symbols to the receiver, tries to persuade the receiver to take an optimal action (from the transmitter's point of view)  by sending a brief summary of the arguments. Finally, it is known that in many practical semantic communication scenarios, e.g., human communication, users with different objectives and background knowledge can coordinate their encoding and decoding strategies to achieve mutual understanding\cite{Cuff2011ImplicitCoordination}. 
This makes strategic communication an ideal tool to investigate semantic communication problems between transmitters and receivers with misaligned distortion measures and unbalanced background knowledge.



In this paper, we investigate strategic semantic communication problems in which the transmitter can obtain a limited number of indirect observations of a semantic information source and can then influence the receiver's decoding process by sending a limited number of messages over a channel with limits on capacity. We focus on the cases that the transmitter and the receiver have different distortion measures and can optimize their encoding and decoding strategies to influence each other. This may correspond to the scenario that a transmitter can have a personal understanding of a semantic message based on its indirect observation results 
and the receiver can then recover the full semantic information based on its own background knowledge as well as  interpretation of the transmitter. We study the impact of strategic decision making on semantic communication and focus particularly on the case of the transmitter committing to an encoding strategy. In this case, the encoder tries to design an optimal strategy to persuade the receiver on a certain feature of the semantic information source based on its indirect observation. We address the following questions: {\em when and how much a transmitter and a receiver can benefit from strategic communication with or without committing to an encoding strategy}.  
Previous results suggested that committing to an encoding strategy can always bring benefit to the encoder\cite{Treust2019Persuasion,Treust2020StrategicCommSideInfo,Akyol2016OnTheRoleStrategicComm}. We, however, prove that there exist cases such that committing to an encoding strategy can in fact harm  the rate distortion performance of the encoder. We therefore provide a sufficient condition under which committing to an encoding strategy can improve the performance of the encoder.
The strategic semantic communication models and solutions considered in this paper have the following unique features:
\begin{itemize}
    \item[(1)] Semantic information source can only be indirectly observed by the encoder and interpreted by the receiver with the assistance of the side information.
    \item[(2)] Transmitter and receiver have different distortion measures and the main objective of the encoder is to influence the receiver's semantic interpretation process instead of maximizing the transporting capacity of information to the receiver.
    \item[(3)] The strategy profiles of encoder and decoder have been derived and compared under three equilibrium solution concepts, including Nash equilibrium, optimal Stackelberg equilibrium, and robust Stackelberg equilibrium. 
\end{itemize}



\vspace{-0.15in}
\section{System Model}
\label{Section_SystemModel}
\vspace{-0.15in}

\begin{figure}[htbp]
	\centering
	\includegraphics[width=3.1 in]{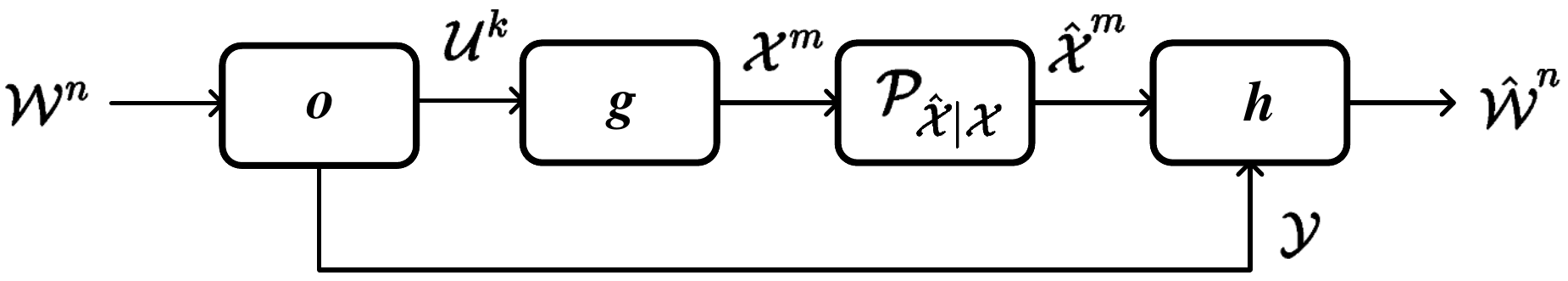}
	\centering
	\caption{\small A semantic communication model consists of a semantic information source $\cW^n$, a signal generator $o$, input of the encoder $\cU^k$, output of the encoder $\cX^m$, channel output $\hcX^m$, side information $\cY$, channel $\cP_{\hcX|\cX}$, output of the decoder $\hcW^n$, encoder $g$ and decoder $h$.}\label{Fig_SemanticCommModel}
\end{figure}

We consider a semantic communication model as illustrated in Fig. \ref{Fig_SemanticCommModel}, where a semantic information source $\cW$ consists of some intrinsic states and features that cannot be directly represented to or observed by the encoder. The encoder can however obtain indirect observations as its input signal source $\cal U$. We use superscript $n$ to denote the length of a sequence of signals generated by the source, e.g., $\cW^n$ denotes a length-$n$ i.i.d. sequence of signals generated by the semantic information source. We consider the case that the encoder can obtain a limited number $k$ of i.i.d. indirect observations ${\cal U}^k$. This may correspond to scenarios where some semantic source may contain rich information and the encoder may have a limited time and resource to conduct sensing and observing.

The decoder has access to a side information $\cY$. 
This side information may correspond to the background knowledge that is helpful for its interpretation of the  semantic  source. For example, in human communication, different combinations and sequences of words may result in different meanings, most of which cannot be directly translated 
by simply combining the meanings of individual words. In this case, the side information, e.g., some common word combinations and relevant background knowledge of the language, known by the intended user will be helpful in interpreting the meaning of the original message. Also, in the context of a smart factory or a smart city, a single instructional message sent by a controller may trigger complex interactions and coordination among a large number of sensors, machines, and systems, each needs to be able to infer the intended meaning and decide on the  responses according to its own functionality and situation. The side information in this case will correspond to the relevant information about the system structure as well as the situation and functions of each system component.  

We consider in this paper semantic communication with a limited channel capacity, and assume that the encoder, after receiving $k$ indirect observations $\cU^k$, needs to send the information via a length-$m$ sequence of messages $\cX^m$. Let $\Delta \left( \cX \right)$ be the set of probability distribution $\cP_\cX$ over $\cX$. We can then define the  encoder's strategy as a mapping $g: \cU^k \rightarrow \Delta \left( \cX^m \right)$ where $\cX^m$ is the channel input. We consider a memoryless channel defined by transition probability  $\cP_{\hcX|\cX}$ where $\hcX^m$ is the channel output. Similarly, the decoder's strategy is defined as a mapping $h: \hcX^m \times \cY \rightarrow \Delta \left( \hcW^n \right)$ where $\hcW^n$ is the output of the decoder, which corresponds to the interpreted  semantic information of the receiver.


In many practical systems, the encoder and the decoder can be associated with different devices and services with different requirements. We therefore consider a strategic communication scenario in which the encoder and the decoder can have different distortion measures 
defined as mappings: $D_{E}: {\cW}\times \cU  \times \cY \times \hcW \rightarrow \mathbb{R}$ and $D_{D}: {\cW}\times \cU \times \cY \times \hcW \rightarrow \mathbb{R}$, respectively.
We focus on  minimizing block-wise distortion between original semantic meaning and the interpreted meaning, given by
\begin{eqnarray}
&& D_{E}\left( \cW, \hcW \right)= \frac{1}{n} \sum\limits^n_{i=1} D_E \left( w_i, {\hw}_i \right),
\label{eq_DE_WW}\\
&& D_{D}\left( \cW, \hcW \right)= \frac{1}{n} \sum\limits^n_{i=1} D_D \left( w_i, {\hw}_i \right).
\label{eq_DD_WW}
\end{eqnarray}


\section{Problem Formulation}
\label{Section_ProblemFormulation}
We consider the following strategic semantic communication scenario: the decoder is aware of the encoding strategy adopted by the encoder and can decide its decoding strategy accordingly by minimizing its own distortion function and the main objective of the encoder is to decide the optimal encoding strategy  (guided by its own distortion measure)  based on the best response of the decoder. As mentioned earlier, simply transporting all the details of its observed information as much as possible may not be the optimal solution and the encoder in this scenario can focus on designing the optimal information structure to influence (e.g., persuade) the decoder to select the optimal strategy in line with his distortion measure.





Before defining the solution profile, we need to first establish links between encoder's and decoder's distortion functions and their strategy profiles by converting (\ref{eq_DE_WW}) and (\ref{eq_DD_WW}) into the following equivalent forms.
\begin{lemma}
Suppose $o$, $g$, $h$, $\cP_{\hcX|\cX}$ are independent. Then, the expected distortion functions of the encoder and the decoder can be rewritten into the following equivalent forms:
\begin{eqnarray}
\lefteqn{D_{E}\left( g, h \right)=} \label{eq_DE_gh}\\
&&\sum_{\cW,\cU,\cY,\hcX,\hcW} \cP_{\cW} \cP_{\cU \cY|\cW} g_{\cX|\cU} \cP_{\hcX|\cX} h_{\hcW|\cY \hcX} D_E \left( \cW, {\hcW} \right), \nonumber\\
\lefteqn{D_{D}\left( g, h \right)=} \label{eq_DD_gh} \\
&&\sum_{\cW,\cU,\cY,\hcX,\hcW} \cP_{\cW} \cP_{\cU \cY|\cW} g_{\cX|\cU} \cP_{\hcX|\cX} h_{\hcW|\cY \hcX} D_D \left( \cW, {\hcW} \right).\nonumber
\end{eqnarray}
\end{lemma}
\begin{IEEEproof}
Following the same line as in \cite{witsenhausen1980indirect}, we have a Markov chain of $\cW,\cU,\cX,\cY,\hcW$. (\ref{eq_DE_gh}) and (\ref{eq_DD_gh}) can then be obtained by substituting $g$ and $h$ into the joint distribution $\cP_{\cW \cU \cY \hcW}$.
\end{IEEEproof}


Let us now define the solution profiles of the strategic semantic communication as follows.


\begin{definition}
\label{Definition_SSE}
In strategic semantic communication, a strategy profile $\langle g^O, h^O \rangle$ is referred to as the {\em optimal Stackelberg equilibrium (OSE)} \cite{bacsar1998dynamic}  if $g^O$ and $h^O$ are given by:
\begin{eqnarray}\
\langle g^O, h^O \rangle &=& \arg \min_{g\in \Delta \left( \cX|\cU \right)} \min_{h \in {\cal H}\left( g \right)} D_E \left( g, h \right)
\label{eq_DefinitionOSE}
\end{eqnarray}
where ${\cal H}\left( g \right) $ is the set of best responses of the decoding strategies under a given encoding strategy $g$, defined as
\begin{eqnarray}
{\cal H}\left( g \right) &=& \left\{h: h=\arg \min_{h\in \Delta \left( \hcW|\cY \hcX \right)} D_D \left( g, h \right)\right\}.
\label{eq_DecodeStrategySet}
\end{eqnarray}
\end{definition}

OSE is the most optimistic solution profile for the encoder because in this case both encoding and decoding strategies have been selected in favor of the encoder.

The other possible strategy profile above is defined as follows:
\begin{definition}
In strategic semantic communication, the strategy profile $\langle g^R, h^R \rangle$ is referred to  as the {\em robust Stackelberg equilibrium (RSE)}  if $g^R$ and $h^R$ are given by:
\begin{eqnarray}
\langle g^R, h^R \rangle &=& \arg \min_{g \in \Delta \left( \cX|\cU \right)} \max_{h \in {\cal H}\left( g \right)} D_E \left( g, h \right)
\label{eq_DefinitionRSE}
\end{eqnarray}
where ${\cal H}\left( g \right)$ is as given in (\ref{eq_DecodeStrategySet}).
\end{definition}

We can observe that RSE provides the worst-case distortion that accrues to the encoder given the response set of the decoder. RSE has been referred to as the optimal robust solution  for the encoder in the literature because it specifies the optimal distortion result the encoder can secure, regardless of which best response chosen by the decoder\cite{Treust2020StrategicCommSideInfo}. 
We can also directly observe that if the encoder and the decoder have the same distortion measures, RSE and OSE result in the same strategy profile.
In this paper, we also compare the above two strategy profiles with the Nash equilibrium (NE) \cite{crawford1982strategic} and we use  $\langle g^N, h^N \rangle$ to denote the strategy profile of NE.





\vspace{-0.1in}
\section{Main Results}
\label{Section_MainResults}

We present the first main result of this paper as follows.


\begin{theorem}
\label{Theorem_DistortionRate}
{For strategic semantic communication, there exists an auxiliary random variable $z \in
\cZ$ and a sufficiently large $m$ such that in OSE and RSE defined in (\ref{eq_DefinitionOSE}) and (\ref{eq_DefinitionRSE}), ${\Delta \left( {\cal X} | {\cal U} \right)}$ is given by: } 
%
\begin{eqnarray}
{\Delta \left( {\cal X} | {\cal U} \right)} &=& \left\{g: \forall {\cP_{\cW \cY \cZ} = g \cP_{\cW} \cP_{\cU \cY | \cW} \cP_{\cZ |\cX}, }\right.\nonumber \\
&&\;\;\; \left. {I \left( \cW, \cU; \cZ | \cY \right)} \le \frac{k}{m} I \left( \cX; \hcX \right)\right\}.
\label{eq_RateConstraint}
\end{eqnarray}
\end{theorem}

\begin{IEEEproof}
To prove the sufficiency of the above result, we follow the same line as in \cite{wyner1978rate,Treust2020StrategicCommSideInfo} to prove that it is possible to construct a joint source and channel coding scheme to achieve the above distortion with all the constraints. Let us  consider an asymptotically optimal source coding scheme concatenated with a reliable channel coding scheme. In particular, we adopt Wyner-Ziv source coding for the source coding scheme and generate a random codebook consisting of $2^{m R_1}$ i.i.d. codewords where $R_1 = I \left(\cW, \cU; \cZ|\cY \right) - \epsilon$. These codewords are then distributed uniformly randomly into $2^{k R_2}$ bins where $R_2 = I \left( \cX; \hcX \right) - \epsilon$. In this way, we compress the source message into $R_1$ bits per symbol with distortion $D_E$ and then map $m R_1$-bit length codeword into a channel coding scheme with $k R_2$ bits. Since $k R_1 \le m R_2$ and also $R_2$ corresponds to the channel capacity, $\cW, \cU, \cZ, \cY, \cX$, and $\hcX$ are jointly typical sequence codes.  
To prove the necessity of Theorem  \ref{Theorem_DistortionRate}, we follow the same line in \cite{Merhav2003JointSourceChannelCode} and introduce an auxiliary random variable $\cZ_i = \langle \cU_i, \hcX_i, \cY^{k}_{i+1}, \cY_{k} \rangle$. We can then prove that $\cY, \cU, \cZ, \cW$ is a Markov chain and exist a joint source and channel encoding scheme $g$ satisfies that
\begin{eqnarray}
I\left(\cW, \cU; \hcX\right) 
&=& I(\cW, \cU, \cY; \hcX) \\
&\ge& I\left(\cW, \cU; \cZ|\cY\right). 
\end{eqnarray}

Similarly, following the same line as in \cite{Merhav2003JointSourceChannelCode,Treust2020StrategicCommSideInfo}, we can prove that for any coding scheme including the proposed joint source and channel encoding, the  mutual information between the coded message and the channel output should not  
exceed the channel capacity $I \left( \cX; \hcX \right)$. We can therefore write $m I\left(\cW, \cU; \hcX\right)$ $\le$ $k I \left( \cX; \hcX \right)$ which concludes the proof.   
\end{IEEEproof}

\noindent{\bf Observation 1:} In the model of strategic semantic communication, the receiver discovers information about the semantic meaning of messages from two sources: (1) the indirect observation obtained by the encoder transmitted through the capacity-limited channel, and (2) the side information available at the receiver. Let us quantify the volumes of information provided by these two sources. The domain of encoding strategy $\cG$ given in (\ref{eq_RateConstraint}) specifies the constraint of the achievable coding rate of the strategic semantic communication. We can rewrite this achievable rate on the left-hand-side of inequality (\ref{eq_RateConstraint}) into the following equivalent forms using the Chain rules,
\begin{eqnarray}
I \left( \cW, \cU; \cZ | \cY \right) &=& I \left( \cW, \cU; \cZ, \cY \right)-I \left( \cW, \cU; \cY \right) \label{eq_RateI}\\
&=& H\left( \cU|\cY \right) + H\left( \cW|\cU, \cY \right) + H\left( \cZ | \cY \right) \nonumber \\
&&\;\;\;\;\;\; - H\left( \cW, \cU, \cZ|\cY \right), \label{eq_RateH}
\end{eqnarray}
where the first term on the right-hand-side of equation (\ref{eq_RateI})  specifies the total amount of uncertainty about the information source $\cW$ and encoder's input $\cU$ that can be reduced at the receiver when being given the channel output $\cZ$ and side information $\cY$. Note that $\cZ$ is an auxiliary random variable introduced in the Wyner-Ziv coding as the output of a ``test channel" for an input $\cX$. $\cZ$ is conditionally independent to $\cY$ given $\cX$. The second term in (\ref{eq_RateI}) specifies the uncertainty about the information source that can be reduced by the side information which is not required to be transported in the channel because the side information is assumed to be already available at the decoder. Equation (\ref{eq_RateH}) quantifies the volume of information, i.e., amount of uncertainty, that can be provided by each individual source. In particular, $H\left( \cU|\cY \right)$ quantifies the information volume of the indirect observation obtained by the encoder. $H\left( \cW|\cU, \cY \right)$ quantifies the ambiguity (uncertainty) about the semantic source when being given the indirect observation of the encoder and side information of the decoder. $H\left( \cW, \cU, \cZ|\cY \right)$ quantifies the uncertainty about the information source and channel output that can be reduced when using the side information at the decoder. In \cite[Theorem 1]{Bao2011SemanticComm}, the authors proved that semantic entropy and observation signal entropy satisfy $H(\cU)=H(\cW)+H(\cU|\cW)-H(\cW|\cU)$ 
where $H(\cU|\cW)$ is defined as the semantic redundancy and $H(\cW|\cU)$ is the semantic ambiguity of coding. Results in (\ref{eq_RateI}) and (\ref{eq_RateH}) can be considered a further step of the result in \cite[Theorem 1]{Bao2011SemanticComm} where we quantify the semantic information source and the input of encoder when being transported into the channel with side information at the decoder.

\noindent{\bf Observation 2:}
One interesting question for strategic semantic communication is whether it is possible for the decoder to simply ``guess" the possible semantic meaning of the source without requiring any information transmitted through the channel? In our previous work\cite{shi2020semantic,XY2022ReasononAir}, a more general scenario for this situation was presented in which the receiver can have access to a knowledge base and can directly infer some of the missing relationships as well as the missing terms or concepts based on this knowledge base. From (\ref{eq_RateI}) and (\ref{eq_RateH}), we can see that the amount of information that can be provided by the side information is given by $H\left( \cW, \cU, \cZ|\cY \right)$ and the total amount of information that can be saved for transmission due to the available side information at the decoder is given by $I \left( \cW, \cU; \cY \right)$.



We have the following results about the strategic semantic communication problems.

\begin{theorem}
\label{Theorem_Stackelberg}

\begin{itemize}
\item[(1)] There exist cases where the encoder's optimal distortion performance achieved by RSE is strictly worse than that achieved by any NE.
\item[(2)] Suppose distortion function $D_E \left(g, h\right)$ is a convex function of $h$ and $\cH(g)$ is a convex domain or $\cH(g)$ is a singleton. The strategic semantic communication then satisfies

    \begin{itemize}
        \item[(2.1)] $D_E \left( g^R, h^R \right) \ge D_E \ge D_E \left( g^O, h^O \right)$;

        \item[(2.2)] There exists at least one strategy profile pair $\langle g^N, h^N \rangle$ such that $D_E \left( g^R, h^R \right) \le D_E \left( g^N, h^N \right)$;

        \item[(2.3)] $D_E \left( g^O, h^O \right) \le D_E \left( g^N, h^N \right)$ for all $\langle g^N, h^N \rangle$. 
    \end{itemize}
\end{itemize}
\end{theorem}
\begin{IEEEproof}
To prove result (1), we only need to provide a contradiction  example to show that RSE can be strictly worse than NE.  
Suppose the strategy profiles of encoder and decoder are probabilities between -1 and 1. 
Decoder's distortion is assumed to be a constant and cannot be affected by $g$ and $h$. This may correspond to the scenario when the decoder has a constant belief about the distortion of the semantic source. Note that in this case any decoding strategy $h$ is a best response. Suppose that the distortion of encoder is given by $D_E = h(h-g)$. Let us first consider NE for this scenario. 
We can show that for any $h>0$, since $D_E$ decreases with $g$, the best response of $g$ is given by $g^N=1$ and we have $\min_g D_E = h^2-h$; for any $h<0$, since $D_E$ increases with $g$, the best response of $g$ is given by $g^N=-1$ and we have $\min_g D_E = h^2+h$; for $h=0$, we have $D_E = 0$. We can therefore claim that, in any NE of our considered case, the distortion of the encoder is always lower than or equal to zero, i.e., $D_E(g^N, h^N)\le 0$. Let us now consider RSE, where the decoder always tries to find the optimal value of $h$ to maximize the encoder's distortion. In particular, we can directly show that when $g\ge0$, the optimal $h$ to maximize $D_E$ is given by $h=-1$ and in this case $\max_h D_E = 1+g \ge 1$; when $g\le0$, the optimal $h$ to maximize $D_E$  is given by $h=1$ and in this case $\max_h D_E = 1-g \ge 1$. We can therefore claim that, in RSE, the optimal distortion of encoder is given by 1, which is much higher than the distortion achieved by any NE. 

Let us now consider result (2). To prove result (2.1), we follow the definitions of $D_E \left( g^O, h^O \right)$ and $D_E \left( g^R, h^R \right)$ and write the following inequality for a given set ${\cal H}\left( g \right)$,
\begin{eqnarray}
\max_{h \in {\cal H}\left( g \right)} D_E \left( g, h \right) \ge \min_{h \in {\cal H}\left( g \right)} D_E \left( g, h \right).
\end{eqnarray}

We can then obtain result (2.1) by taking minimization over both sides of the above inequality.


To prove result (2.2), let us first define the sets of best responses of encoder and decoder as $\cG (h)$ and $\cH (g)$, respectively. We can then define a worst-case NE strategy profile of encoder and decoder as follows:
\begin{eqnarray}
\langle {\tilde g}^{N}, {\tilde h}^{N*} \rangle = \max_{g \in \cG (h)} \max_{h \in \cH (h)} D_E \left( g, h \right).
\end{eqnarray}

We can directly see that $D_E \left( g^R, h^R \right) \le D_E \left( {\tilde g}^{N}, {\tilde h}^{N*} \right)$ and $D_E \left( g^O, h^O \right) \le D_E \left( {\tilde g}^{N}, {\tilde h}^{N*} \right)$ for all the strategy profile pairs $\langle g^R, h^R \rangle$ and $\langle g^O, h^O \rangle$.


To prove result (2.3), let us first prove that the domain $\cal G$ of $g$ is a convex closure. From (\ref{eq_RateConstraint}), we can observe that since $I(\cX; \hcX)$ is the channel capacity which can be assumed to be a constant in our setting, the domain of $g$ can be obtained by adjoining all linear combinations of points satisfying the constraints in (\ref{eq_RateConstraint}) which is a convex closure.  

For the rest of the proof, we only need to show that there does not exist an NE $\langle g^N, h^N \rangle$ that can achieve a lower distortion $D_E (g, h)$ than an OSE. We can prove this by contradiction. Let us assume that there exists an NE $\langle {\hat g}^N, {\hat h}^N \rangle$ that satisfies $D_E \left({\hat g}^N, {\hat h}^N \right) < D_E \left(g^O, h^O \right)$. If this is the case, we have
\begin{eqnarray}
&&\min_h D_E (g, h) < \min_g \min_{h\in \cH (g)} D_E (g, h) \; \mbox{or} \nonumber \\
&&\min_g D_E (g, h) < \min_g \min_{h\in \cH (g)} D_E (g, h)\; \mbox{for any} \langle g, h \rangle \nonumber 
\end{eqnarray}
which cannot be true. This concludes our proof.
\end{IEEEproof}

\noindent{\bf Observation 3:} In the previous literature, RSE has been commonly referred to as the optimal robust solution for the encoder and has been considered in some previous works as a promising solution concept for strategic communication\cite{Akyol2016OnTheRoleStrategicComm,Treust2020StrategicCommSideInfo,Treust2021ISITStrategicCommSideInfo}. Our result (1) however suggests that it is possible that RSE is strictly worse than any NE.  In this case, a better strategy for the encoder is to avoid RSE by enabling simultaneous decision making or allowing leadership swap between encoder and decoder\cite{bacsar1998dynamic}. 

\noindent{\bf Observation 4:} Result (2.1) in Theorem \ref{Theorem_Stackelberg} provides an upper and a lower bound on the distortion that can be expected by the encoder. It also provides a sufficient condition under which committing to an encoding strategy can always reduce the distortion for the encoder. 
We can observe that in a special case such that distortion functions of encoder and decoder have the same monotonic trend over $g$ and $h$, then both encoder and decoder will have the incentive to select the optimal strategy profile to minimize the distortion. The upper and lower bounds of $D_E$ become equal when both encoder and decoder have the same distortion measure, i.e., $D_E (g, h) = D_D (g, h)$. 


\begin{figure}[htbp]
	\centering
	\begin{minipage}{0.45\linewidth}
	\footnotesize
	\centering
	\captionof{table}{\footnotesize{Distortion Matrix}}
	\setlength{\tabcolsep}{1.5 pt}{
	\begin{tabular}{|c|c|c|c|}
    \hline
    &  $h_{0}$ & $h_{1}$ & $h_{2}$\\
    \hline
    $g_{0}$ & (0,0) & ($\alpha$,$\beta$+1.2) &(7,8)\\
    \hline
    $g_{1}$ & ($\alpha$,$\beta$) & (0,1.2) & (6,7)\\
    \hline
    $g_{2}$ & (7,7) & (6,7.2) & (0,1) \\
    \hline
    \end{tabular}}
	\end{minipage}
	\hfill
	\begin{minipage}{0.45\linewidth}
	\centering
	\includegraphics[width=1.5 in]{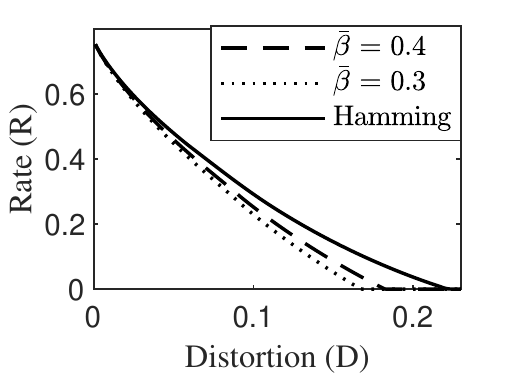}
	\vspace{-0.1in}
	\footnotesize
	\caption{\footnotesize{R-D Function}}
	\label{Fig_RateDistortionFunction}
	\end{minipage}
	\vspace{-0.2in}
\end{figure}

\begin{figure}[htbp]
	\centering
	\begin{minipage}{0.42\linewidth}
	\centering
	\includegraphics[width=\textwidth]{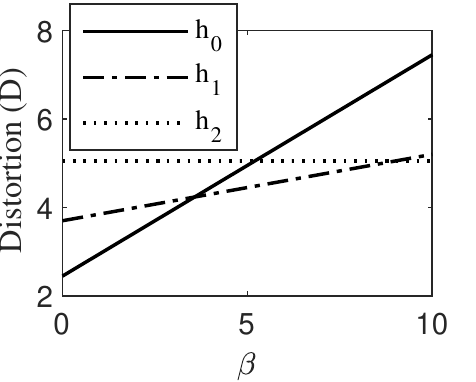}
	\vspace{-0.2in}
	\footnotesize
	\caption{\footnotesize{Distortion of Decoder}}
	\label{Fig_DecoderDistortion}
	\end{minipage}
	\centering
    \hfill
	\begin{minipage}{0.45\linewidth}
	\centering
	\includegraphics[width=\textwidth]{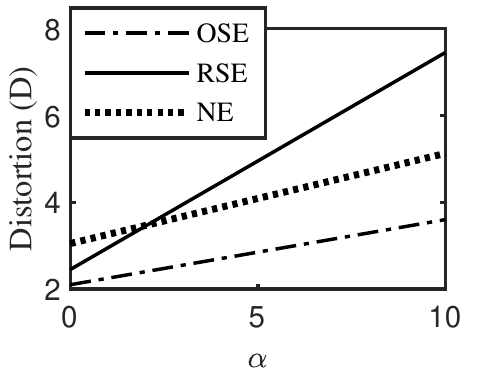}
	\caption{\footnotesize{Distortion of Encoder. }}
	\label{Fig_EncoderDistortion}
	\end{minipage}
	\centering
	\vspace{-0.3in}
\end{figure}

\section{Example with Multinoulli Semantic Source}
\label{Multinoulli}

We consider a semantic communication example in which the semantic information source $\cW$ is a Multinoulli random variable uniformly randomly drawn from a dictionary-based knowledge dataset, WordNet. We consider distortion function specified by the meaning dissimilarity between different semantic symbols as defined in the WordNet dataset. In Table I, we present the encoder and decoder's semantic distortion matrix of three different symbols (words) where first and last symbols (labeled as 0 and 2) may correspond to ``Nickel" and ``Gold" and symbol 1 can be selected from a set of possible symbols with the semantic distances to symbol 0 specified by values of $\alpha$ and $\beta$ for encoder and decoder, respectively, e.g., if symbol 1 is ``Coin", $\alpha=\beta=1$. We also introduce some bias between distortions at the encoder and decoder to reflect the difference between the personal preference or backgrounds of the transmitter and the receiver. We present the rate distortion function with different values of $\beta$ normalized to compared with Hamming distance-based distortion measure in Fig. \ref{Fig_RateDistortionFunction}. We observe that since semantic sources may consist of symbols with relatively less distortion when being incorrectly decoded, it may result in less distortion than the Hamming distance which is equal between any different symbols. In Fig. \ref{Fig_DecoderDistortion} and \ref{Fig_EncoderDistortion}, we present the distortion of decoder and encoder under different values of $\alpha$ and $\beta$. We can observe that different decoding strategies may have different distortions. Also, in some cases, RSE can be strictly worse than both NE and OSE. 


\vspace{-0.2in}
\section{Conclusions}
\label{Section_Conclusion}
We have investigated the fundamental limit of distortion rate for strategic semantic communication problems. 
We focus on the case when the transmitter commits to an encoding strategy. 
To investigate when and how much a transmitter and a receiver can benefit from strategic communication with or without committing to an encoding strategy, we have studied three types of equilibrium solutions: OSE, RSE, and NE. The optimal encoding and decoding strategy profiles have been derived.
We have observed that committing to an encoding strategy cannot always achieve distortion reduction at the encoder. We have accordingly proposed a feasible condition under which committing to an encoding strategy can always improve distortion performance.

\bibliographystyle{IEEEtran}
\bibliography{ref.bib}

\begin{thebibliography}{10}
\providecommand{\url}[1]{#1}
\csname url@samestyle\endcsname
\providecommand{\newblock}{\relax}
\providecommand{\bibinfo}[2]{#2}
\providecommand{\BIBentrySTDinterwordspacing}{\spaceskip=0pt\relax}
\providecommand{\BIBentryALTinterwordstretchfactor}{4}
\providecommand{\BIBentryALTinterwordspacing}{\spaceskip=\fontdimen2\font plus
\BIBentryALTinterwordstretchfactor\fontdimen3\font minus
  \fontdimen4\font\relax}
\providecommand{\BIBforeignlanguage}[2]{{%
\expandafter\ifx\csname l@#1\endcsname\relax
\typeout{** WARNING: IEEEtran.bst: No hyphenation pattern has been}%
\typeout{** loaded for the language `#1'. Using the pattern for}%
\typeout{** the default language instead.}%
\else
\language=\csname l@#1\endcsname
\fi
#2}}
\providecommand{\BIBdecl}{\relax}
\BIBdecl

\bibitem{shannon1948mathematical}
C.~E. Shannon, ``A mathematical theory of communication,'' \emph{The Bell
  system technical journal}, vol.~27, no.~3, pp. 379--423, 1948.

\bibitem{weaver1949recent}
W.~Weaver, ``Recent contributions to the mathematical theory of
  communication,'' \emph{ETC: a review of general semantics}, pp. 261--281,
  1949.

\bibitem{Carnap1952SemanticInfo}
R.~Carnap and Y.~Bar-Hillel, ``An outline of a theory of semantic
  information,'' \emph{Technical report, Massachusetts Institute of Technology.
  Research Laboratory of Electronics}, October 1952.

\bibitem{Bao2011SemanticComm}
J.~Bao, P.~Basu, M.~Dean, C.~Partridge, A.~Swami, W.~Leland, and J.~A. Hendler,
  ``Towards a theory of semantic communication,'' in \emph{IEEE Network Science
  Workshop}, West Point, NY, Jun. 2011, pp. 110--117.

\bibitem{Liu2021SemanticInformation}
J.~Liu, W.~Zhang, and H.~V. Poor, ``A rate-distortion framework for
  characterizing semantic information,'' in \emph{IEEE ISIT}, Melbourne,
  Australia, Jul. 2021, pp. 2894--2899.

\bibitem{shi2020semantic}
G.~Shi, Y.~Xiao, Y.~Li, and X.~Xie, ``From semantic communication to
  semantic-aware networking: Model, architecture, and open problems,''
  \emph{IEEE Commun. Magazine}, vol.~59, no.~8, pp. 44--50, Aug. 2021.

\bibitem{Juba2008UniverSemanticComm}
B.~Juba and M.~Sudan, ``Universal semantic communication i,'' in
  \emph{Proceedings of the ACM Symposium on Theory of Computing}, Victoria,
  British Columbia, Canada, May. 2008.

\bibitem{Floridi2010InformationBook}
L.~Floridi, \emph{Information: A very short introduction}.\hskip 1em plus 0.5em
  minus 0.4em\relax Oxford University Press, 2010.

\bibitem{Floridi2012SemanticInfo}
------, ``Semantic information and the network theory of account,''
  \emph{Synthese}, vol. 184, no.~3, pp. 431--454, Feb. 2012.

\bibitem{StanfordEncyclopedia2022SemanticInfo}
S.~Sequoiah-Grayson and L.~Floridi, ``{Semantic Conceptions of Information},''
  in \emph{The {Stanford} Encyclopedia of Philosophy}, {S}pring 2022~ed., E.~N.
  Zalta, Ed.\hskip 1em plus 0.5em minus 0.4em\relax Metaphysics Research Lab,
  Stanford University, Jan. 2022.

\bibitem{Akyol2015StrategicCompression}
E.~Akyol, C.~Langbort, and T.~Ba{\c{s}}ar, ``Strategic compression and
  transmission of information,'' in \emph{IEEE ITW}, Jeju, Korea, Oct. 2015,
  pp. 219--223.

\bibitem{Akyol2016OnTheRoleStrategicComm}
------, ``On the role of side information in strategic communication,'' in
  \emph{IEEE ISIT}, Barcelona, Spain, Jul. 2016, pp. 1626--1630.

\bibitem{Akyol2017StrategicComm}
------, ``Information-theoretic approach to strategic communication as a
  hierarchical game,'' \emph{Proceedings of the IEEE}, vol. 105, no.~2, pp.
  205--218, Feb. 2017.

\bibitem{Nisan2007AlgorithmGameTheory}
N.~Nisan, T.~Roughgarden, E.~Tardos, and V.~V. Vazirani, \emph{Algorithmic Game
  Theory}.\hskip 1em plus 0.5em minus 0.4em\relax Cambridge University Press,
  2007.

\bibitem{Treust2019Persuasion}
M.~Le~Treust and T.~Tomala, ``Persuasion with limited communication capacity,''
  \emph{Journal of Economic Theory}, vol. 184, p. 104940, 2019.

\bibitem{Treust2020StrategicCommSideInfo}
------, ``Strategic communication with side information at the decoder,''
  arXiv:1911.04950, Sep. 2020.

\bibitem{Treust2021ISITStrategicCommSideInfo}
------, ``Strategic communication with decoder side information,'' in
  \emph{IEEE ISIT}, Melbourne, Australia, Jul. 2021, pp. 2696--2701.

\bibitem{kamenica2011bayesian}
E.~Kamenica and M.~Gentzkow, ``Bayesian persuasion,'' \emph{American Economic
  Review}, vol. 101, no.~6, pp. 2590--2615, 2011.

\bibitem{Cuff2011ImplicitCoordination}
P.~Cuff and L.~Zhao, ``Coordination using implicit communication,'' in
  \emph{IEEE ITW}, 2011, pp. 467--471.

\bibitem{witsenhausen1980indirect}
H.~Witsenhausen, ``Indirect rate distortion problems,'' \emph{IEEE Transactions
  on Information Theory}, vol.~26, no.~5, pp. 518--521, 1980.

\bibitem{bacsar1998dynamic}
T.~Ba{\c{s}}ar and G.~J. Olsder, \emph{Dynamic noncooperative game
  theory}.\hskip 1em plus 0.5em minus 0.4em\relax SIAM, 1998.

\bibitem{crawford1982strategic}
V.~P. Crawford and J.~Sobel, ``Strategic information transmission,''
  \emph{Econometrica: Journal of the Econometric Society}, pp. 1431--1451,
  1982.

\bibitem{wyner1978rate}
A.~D. Wyner, ``The rate-distortion function for source coding with side
  information at the decoder$\backslash$3-ii: General sources,''
  \emph{Information and control}, vol.~38, no.~1, pp. 60--80, 1978.

\bibitem{Merhav2003JointSourceChannelCode}
N.~Merhav and S.~Shamai, ``On joint source-channel coding for the wyner-ziv
  source and the gel'fand-pinsker channel,'' \emph{IEEE Transactions on
  Information Theory}, vol.~49, no.~11, pp. 2844--2855, 2003.

\bibitem{XY2022ReasononAir}
Y.~Xiao, Y.~Li, G.~Shi, and H.~V. Poor, ``Reasoning on the air: An implicit
  semantic communication architecture,'' in \emph{Proc. of the IEEE ICC
  Workshop on Data Driven Intelligence for Networks and Systems}, Seoul, South
  Korea, May 2022.

\end{thebibliography}

\end{document}